\documentclass[a4paper,12pt]{article}
\usepackage{graphicx,rotating,amsmath,amssymb}
\usepackage{hyperref}\usepackage{slashed,cite}
\pdfoutput=1
\hypersetup{colorlinks,bookmarksopen,bookmarksnumbered,
linkcolor=blus,pdfstartview=FitH,urlcolor=rossos,citecolor=verde}

\newcommand{\med}[1]{\langle #1 \rangle}

\newcommand{\fig}[1]{~\ref{fig:#1}}

\usepackage{soul}
\usepackage{multicol}
\usepackage{color}
\definecolor{rosso}{cmyk}{0,1,1,0.4}
\definecolor{rossos}{cmyk}{0,1,1,0.55}
\definecolor{rossoc}{cmyk}{0,1,1,0.2}
\definecolor{blu}{cmyk}{1,1,0,0.3}
\definecolor{blus}{cmyk}{1,1,0,0.6}
\definecolor{bluc}{cmyk}{1,1,0,0.1}
\definecolor{verde}{cmyk}{0.92,0,0.59,0.25}
\definecolor{verdec}{cmyk}{0.92,0,0.59,0.15}
\definecolor{verdes}{cmyk}{0.92,0,0.59,0.4}
\newcommand{\eq}[1]{~{\rm (\ref{eq:#1})}}

\newcommand{\GeV}{\,{\rm GeV}}
\newcommand{\TeV}{\,{\rm TeV}}

\def\circa#1{\,\raise.3ex\hbox{$#1$\kern-.75em\lower1ex\hbox{$\sim$}}\,}

\newcommand{\beq}{\begin{equation}}
\newcommand{\eeq}{\end{equation}}

\newcommand{\bea}{\begin{eqnarray}}
\newcommand{\eea}{\end{eqnarray}}
\newcommand{\be}{\begin{equation}}
\newcommand{\ee}{\end{equation}}
\font\tenrsfs=rsfs10 at 12pt
\font\sevenrsfs=rsfs7
\font\fiversfs=rsfs5
\newfam\rsfsfam
\textfont\rsfsfam=\tenrsfs
\scriptfont\rsfsfam=\sevenrsfs
\scriptscriptfont\rsfsfam=\fiversfs
\def\mathscr#1{{\fam\rsfsfam\relax#1}}

\def\circa#1{\,\raise.3ex\hbox{$#1$\kern-.75em\lower1ex\hbox{$\sim$}}\,}
\makeatletter

\def\hhref#1{\href{http://arxiv.org/abs/#1}{arXiv:#1}} % in bibliography

\def\hhref#1{\href{http://arxiv.org/abs/#1}{arXiv:#1}}
\usepackage{xstring}
\newcommand{\hhrefq}[1]{\IfSubStr{#1}{:}{\href{http://inspirehep.net/search?ln=en&ln=en&p=#1&of=hb&action_search=Search&sf=&so=d&rm=&rg=25&sc=0}{InSpire:#1}}{\hhref{#1}}}

\def\art{\@ifnextchar[{\eart}{\oart}}
\def\eart[#1]#2#3#4#5#6{{\rm #2}, {\em #3 \bf #4} {\rm (#6) #5} ({\em #1})}
\def\article{\@ifnextchar[{\earticle}{\oarticle}}
\def\oarticle#1#2#3#4#5#6{{\rm #1}, {\ital `#6'}, {\rm #2 #3 (#5) #4}}
\def\earticle[#1]#2#3#4#5#6#7{{\rm #2}, {\ital `#7'}, {\rm #3 #4 (#6) #5}  [\hhrefq{#1}]}
\def\hepart[#1]#2{{\rm #2, \sl#1}}
\def\heparticle[#1]#2#3{#2, {\ital `#3'} [\hhrefq{#1}]}
\newcommand{\doi}[1]{\href{http://dx.doi.org/#1}{[link]}}
\newcommand{\hhrefqq}[1]{\IfBeginWith{#1}{10.}{\href{https://doi.org/#1}{doi:#1}}{\hhrefq{#1}}}

\renewenvironment{thebibliography}[1]
{\begin{multicols}{2}[\section*{\refname}]%
		\@mkboth{\MakeUppercase\refname}{\MakeUppercase\refname}%
		\list{\@biblabel{\@arabic\c@enumiv}}%
		{\settowidth\labelwidth{\@biblabel{#1}}%
			\leftmargin\labelwidth
			\advance\leftmargin\labelsep
			\@openbib@code
			\usecounter{enumiv}%
			\let\p@enumiv\@empty
			\renewcommand\theenumiv{\@arabic\c@enumiv}}%
		\sloppy
		\clubpenalty4000
		\@clubpenalty \clubpenalty
		\widowpenalty4000%
		\sfcode`\.\@m}
	{\renewcommand{\@noitemerr}
		{\@latex@warning{Empty `thebibliography' environment}}%
		\endlist\end{multicols}}

\font\ital=cmu10

\newcounter{alphaequation}[equation]
\def\thealphaequation{\theequation\hbox to
0.6em{\hfil\alph{alphaequation}\hfil}}
\def\eqnsystem#1{
\def\@eqnnum{{\rm (\thealphaequation)}}
\def\@@eqncr{\let\@tempa\relax \ifcase\@eqcnt \def\@tempa{& & &} \or
  \def\@tempa{& &}\or \def\@tempa{&}\fi\@tempa
  \if@eqnsw\@eqnnum\refstepcounter{alphaequation}\fi
\global\@eqnswtrue\global\@eqcnt=0\cr}
\refstepcounter{equation} \let\@currentlabel\theequation \def\@tempb{#1}
\ifx\@tempb\empty\else\label{#1}\fi
\refstepcounter{alphaequation}
\let\@currentlabel\thealphaequation
\global\@eqnswtrue\global\@eqcnt=0 \tabskip\@centering\let\\=\@eqncr
$$\halign to \displaywidth\bgroup \@eqnsel\hskip\@centering
$\displaystyle\tabskip\z@{##}$&\global\@eqcnt\@ne
\hskip2\arraycolsep\hfil${##}$\hfil& \global\@eqcnt\tw@\hskip2\arraycolsep
$\displaystyle\tabskip\z@{##}$\hfil
\tabskip\@centering&\llap{##}\tabskip\z@\cr}
\def\endeqnsystem{\@@eqncr\egroup$$\global\@ignoretrue} \makeatother

\newcommand{\U}{\,{\rm U}}

\newcommand{\dd}{\mathrm d}

\allowdisplaybreaks[1]
\definecolor{ccmag}{cmyk}{0,1,0,0.2}
\definecolor{ccpur}{cmyk}{0.6,0.95,0,0.3}

\usepackage{array}
\newcolumntype{L}[1]{>{\raggedright\arraybackslash}p{#1}}

\begin{document}

\begin{center}
{\LARGE \bf \color{rossos}
Cosmological vacuum selection in dynamical symmetry breaking at zero vacuum energy}\\
\bigskip\bigskip
{\large\bf Mateusz Duch} {\large and} {\large\bf Alessandro Strumia}\\[2ex]
{\it Dipartimento di Fisica, Universit{\`a} di Pisa, Italia}\\

\bigskip\bigskip

{\large\bf\color{blus} Abstract}
\begin{quote}\large
Coleman--Weinberg models invoke classical scale invariance to forbid scalar masses,
but then add a dimensionful constant to cancel the vacuum energy of the broken vacuum.
This inconsistency is avoided by special quantum potentials,
whose symmetry-breaking vacuum is degenerate with the symmetric one.
We compute which vacuum is selected by the thermal history of the universe.
While a neutral thermal plasma favours the symmetric phase, a sufficiently large conserved charge asymmetry supports a rotating scalar condensate that evolves toward the broken vacuum as the universe cools, and survives fragmentation.
\end{quote}
\thispagestyle{empty}
\end{center}
\tableofcontents
\setcounter{footnote}{0}

\newpage

\section{Introduction}
Classically scale-invariant theories allow for dynamical symmetry breaking near the scale $w$ at which the running quartic coupling $\lambda$ of a scalar $S$ crosses zero and becomes negative~\cite{Coleman:1973jx,Gildener:1976ih}. 
 Additional interactions can generate a beta function $\beta_\lambda$, giving approximately $\lambda(S)\sim\beta_\lambda\ln(S/w)$. The resulting Coleman--Weinberg potential takes the form $V(S)\sim\beta_\lambda S^4\ln(S/w)$,
 so that $S=0$ is unstable.
In the hot early universe, $V$ acquires a positive thermal mass term $\sim T^2S^2$, which
adds a potential barrier and initially traps the field at $S=0$. 
If the couplings are sufficiently large for tunnelling to complete, the universe tunnels to the deeper vacuum at $S\sim w$. 
The transition tends to be strongly first-order and to produce observable cosmological signals,
see e.g.~\cite{0902.4050,1104.4791,1306.2329,1310.4423,1604.05035,1704.01034,1809.11129,2506.16387,2607.15279}.

However, observations require the present vacuum energy to be nearly zero. A constant $V_\Lambda\sim\beta_\lambda w^4$ must therefore be added to the potential and tuned to cancel its value at the broken minimum. This introduces the explicit dimensionful parameter $V_\Lambda$, in tension with the original appeal to classical scale invariance as the reason for omitting a scalar mass term. This issue is usually set aside. 
A possible justification is that $V_\Lambda$ affects only gravity, which in any case involves the Planck mass;
it may therefore be reasonable to assume classical scale invariance only in the non-gravitational sector.

Here we explore whether this issue can be addressed without adding a constant to the potential. Since the symmetric vacuum has $V(0)=0$, requiring the broken vacuum to have exactly vanishing energy implies two degenerate vacua: one at $S=0$, and another at $S=w$ tuned to have $V(w)=0$. 
This kind of potential can again be generated by quantum effects,
provided that the running is tuned such that $\lambda$ and $\beta_\lambda$ vanish at the same scale $w$.
This nearly happens in the Standard Model, 
where an extra unphysical minimum can arise around $w\sim 10^{17}\,\GeV$~\cite{hep-ph/9511371,1307.3536}.
The resulting potential is approximately $V(S)\propto S^4\ln^2(S/w)$, and features a barrier separating $S=0$ from $S=w$.
Previous work~\cite{1403.4226} assumed that the universe sits in the broken minimum after inflation, and
that the potential barrier is large enough for the subsequent thermal phase not to destabilise it.
We go further and ask whether, in these models, the broken vacuum can be dynamically selected
by the subsequent cosmological thermal phase.
The two vacua are degenerate but not equivalent, so cosmological dynamics can make a choice.

In section~\ref{minimalmodel} we show that thermal effects favour $S=0$.
This is non-trivial since, in a general theory, 
mixed scalar quartics can give negative squared thermal masses and favour symmetry breaking \cite{Weinberg:1974hy}.
This is a multi-field phenomenon, while Coleman--Weinberg potentials are nearly flat only along a one-dimensional ray in field space. Along this ray all particles are massless at the origin and become massive away from it, so thermal masses are positive. Along other directions the tree-level quartics dominate, and any thermal minimum returns to the origin as the universe cools.

In section~\ref{charge} we show that this conclusion can be evaded in the presence of a conserved charge which is stored in the rotating condensate of a complex scalar $S$.
We verify that the broken vacuum can be dynamically selected and can resist fragmentation.
Conclusions are given in section~\ref{concl}.

\begin{figure}[t]
$$\includegraphics[width=0.6\textwidth]{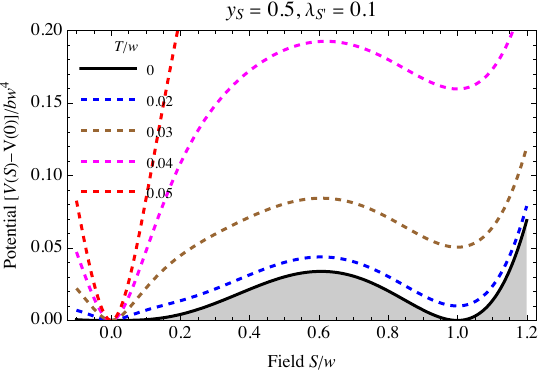}$$
\begin{center}
\caption{\em\label{fig:HonestColemanWeinbergV} Thermal potential in the model of eq.\eq{realmodel}, for the indicated values of the parameters.}
\end{center}
\end{figure}

\section{Vacuum selection by thermal effects}\label{minimalmodel}
To see the general point, we start considering a minimal model containing two real singlets $S,{S'}$ and one Weyl fermion $\psi$, with canonical
kinetic terms and the following interactions
\begin{equation}
 V=\frac{\lambda_S}{4}S^4+\frac{\lambda_{SS'}}{4}S^2{S'}^2
       +\frac{\lambda_{S'}}{4}{S'}^4,
 \qquad
  \mathscr{L}_{\rm Yuk}=-\frac12 y_{S}S\psi\psi+\mathrm{h.c.}
 \label{eq:realmodel}
\end{equation}
For simplicity we imposed a ${S'}\to-{S'}$ symmetry, which forbids ${S'}\psi\psi$, $S^3{S'}$ and $S{S'}^3$.
Couplings to the SM are neglected here.
With $\beta_g=\dd g/\dd\ln\bar\mu$, the one-loop RGEs are
\begin{eqnsystem}{sys:RGE2}
\label{eq:realrge}
(4\pi)^2\beta_{\lambda_S}&=&18\lambda_S^2+\tfrac12\lambda_{SS'}^2
                      +4\lambda_S y_{S}^2-4y_{S}^4,\\
(4\pi)^2\beta_{\lambda_{S'}}&=&18\lambda_{S'}^2+\tfrac12\lambda_{SS'}^2,\\
(4\pi)^2\beta_{\lambda_{SS'}}&=&4\lambda_{SS'}^2+6\lambda_{SS'}(\lambda_S+\lambda_{S'})+2\lambda_{SS'} y_{S}^2,\\
(4\pi)^2\beta_{y_S}&=&4y_{S}^3.
\end{eqnsystem}
These RGEs allow the needed running, thanks to the competing effects of the Yukawa and of the scalar quartics.
Depending on the values of the running couplings, two qualitatively different patterns of symmetry breaking can happen
(breaking patterns for general scalar potentials are discussed in \cite{1603.02680}):
\begin{itemize}
\item[1)] If the running crosses $\lambda_{S}=0$ while $\lambda_{S'}, \lambda_{SS'}>0$,
one has a nearly-flat potential along $S'=0$.
The quantum potential at zero temperature
\beq \label{eq:VSlog2}
V_0(S,0) \approx \lambda_{S}(S) \frac{S^4}{4}\approx bS^4\ln^2\frac{S}{w}\eeq 
has a
 minimum at $S=w$ and $S'=0$ with tuned vanishing cosmological constant  if
\begin{equation}
 \lambda_S(w)=\beta_{\lambda_{S}}(w)=0,
 \qquad \beta'_{\lambda_{S}}(w)>0.
 \label{eq:vacuumconditions}
\end{equation}
The equations are solved by
\begin{equation}
y_{S}^2(w) =\frac{ \lambda_{SS'}(w)}{2\sqrt2},\qquad
 b=\frac{\beta'_{\lambda_S}(w)}8
 =\frac{\lambda_{SS'}^2[6\lambda_{S'}+(4-3/\sqrt2)\lambda_{SS'}]}
             {8(4\pi)^4}>0.
 \label{eq:realb}
\end{equation}

At finite temperature $T$, the potential acquires thermal corrections.
Around the origin, they are encoded in thermal masses 
\begin{equation}
M^2_S=\frac{T^2}{24}(6\lambda_S+\lambda_{SS'}+2y_{S}^2),\qquad
M^2_{S'}=\frac{T^2}{24}(6\lambda_{S'}+\lambda_{SS'}),\qquad M_{SS'}^2 = 0.
 \label{eq:thermalreal}
\end{equation}
Since $M^2_S>0$, thermal corrections dynamically select the minimum at $S=0$, 
as confirmed by computing the full effective potential, plotted in fig.\fig{HonestColemanWeinbergV}
along $S'=0$.

\item[2)]
If the running crosses $\lambda_{SS'}=-2\sqrt{\lambda_S \lambda_{S'}}<0$ while $\lambda_S, \lambda_{S'}>0$
the nearly-flat $V\simeq 0$ line is at $S/S'=(\lambda_{S'}/\lambda_S)^{1/4}$.
One can naively hope that the negative mixed quartic $\lambda_{SS'}$ generates 
a large enough negative contribution to the squared thermal masses to destabilise the origin.
So we proceed with the detailed computation.
The potential along the flat direction $\phi^2 =S^2 + S'^2$ 
can again be approximated as $V_0 = b \phi^4 \ln^2(\phi/w)$,
after tuning $y_S$ to get a vanishing cosmological constant 
\beq 
y_S^2(w)=\sqrt2\Big(\lambda_S-\frac{\lambda_{SS'}}{2}\Big),
\qquad
b=\frac{\lambda_{SS'}^2\big[(5-2\sqrt2)\lambda_S+(5+3/\sqrt2)\lambda_{SS'}+3\lambda_{S'}\big]}{(4\pi)^4}.
\label{eq:yScase2}
\eeq
The factor $b$ (being two-loop suppressed) receives a novel contribution from the RGE rotation of the flat direction
and is positive if $\lambda_S/\lambda_{S'}$ is either large enough or small enough.
The thermal masses at the origin are still given by eq.\eq{thermalreal}, 
and $M_{S'}^2$ is negative in the large $\lambda_S/\lambda_{S'}$ branch.
However, the thermal mass along the flat direction is positive,
\beq M_\phi^2 = -\Big(1+\frac1{\sqrt2}\Big)\frac{\lambda_{SS'}}{12}\,T^2>0.\eeq
So the thermal vacuum can be away from the origin, but approaches it as $T\to 0$.
The broken phase is not dynamically selected by the thermal history.
\end{itemize}
Other attempts with more complicated models similarly fail.
% \footnote{A different conclusion was reached in~\cite{2211.09045}, where (section~4.2) the contribution of the thermalised Higgs to the dilaton thermal mass is omitted, on the grounds that the dilaton is not thermalised. Including it, the thermal minimum returns to the origin as the universe cools, in agreement with our results.}
The general reason is the following.

\subsection{Why a neutral thermal plasma favours the origin}\label{sec:thermal}

For any tree-level flat ray $\phi$, the field-dependent squared masses scale as $m_i^2(\phi)=c_i\phi^2$. If the ray is stable in all transverse scalar directions, the scalar mass eigenvalues have $c_i\geq 0$. In case of fermions and gauge-bosons, $c_i\geq 0$ by construction. It follows that the thermal mass is non-negative along the flat direction
\begin{equation}
M_\phi^2=\frac{T^2}{12}
 \left[\sum_b n_bc_b+\frac12\sum_f n_fc_f\right]\geq0.
\end{equation}
More in general, the one-loop thermal potential is written in terms of thermal functions $J_{B,F}(m^2/T^2)$
which increase with $m^2$.
Consequently each stable free-particle contribution satisfies $V_T(\phi)-V_T(0)\geq0$.
The vacuum at the origin is thermally selected because it is the unique special point 
where all masses vanish (in the absence of other mass-generating sectors).
In more complicated models, extra fermions can be chosen such that some combination of their masses cancels along the ray: their contribution can then vanish, but cannot become negative.

The origin can instead be destabilised along a direction oblique to the flat ray, 
where the squared thermal mass can be negative, as in case 2) above.
Along such a direction the tree-level quartic is positive and, unless it is loop-suppressed
(comparable to $\lambda_{SS'}^2/(4\pi)^2$), it dominates over radiative corrections.
The negative thermal mass term is then balanced by the tree-level quartic, producing a thermal minimum
at field values of order $T$ (up to powers of coupling ratios).
This minimum scales as $T$ and smoothly returns to the origin as $T\to 0$: 
the vacuum at $w$ becomes a local minimum only at $T\ll w$,
when the thermal minimum already sits at field values $\sim T\ll w$, on the origin's side of the barrier.

In models of Coleman--Weinberg electroweak symmetry breaking
the $\med{S}=0$ minimum gets lowered by the QCD vacuum energy from the gluon condensate, and by the
QCD Higgs-top tadpole~\cite{1704.04955,1709.10322}.
In our context this does not destabilise $S=0$ towards the $\med{S}\neq 0$ minimum.
A contribution which merely lowers the energy of the
$S=0$ phase does not drive a transition to the zero-energy vacuum at $w$.

We neglected the contribution to the thermal potential from the light ray mode $\phi$ itself,
because its potential is two-loop suppressed.
Its high-temperature contribution starts as $T^2V_0''(\phi)/24$ and $V_0''<0$ around the top of the barrier,
possibly leading to qualitatively new effects after resummation.
This can only become relevant for non-perturbative couplings, so we avoid exploring this direction.

\section{Vacuum selection by a particle asymmetry}\label{charge}
We now investigate whether a particle asymmetry can lead to the cosmological selection of the broken vacuum in our scenario. Conserved charges are known to induce Bose--Einstein condensation and spontaneous symmetry breaking at finite temperature~\cite{Kapusta:1981aa,Haber:1981ts}.
They have also been shown to prevent electroweak symmetry restoration at high temperature~\cite{2210.05680}.
To accommodate a conserved charge, we promote $S$ to a complex scalar carrying unit charge under an exact global `dark' $\U(1)$ symmetry $S(x) =s(x)\, e^{i\theta(x)}/\sqrt2$. We keep ${S'}$ neutral and assign charge $-1/2$ to the Weyl fermion. Together with classical scale invariance, the global $\U(1)$ symmetry implies an accidental \(S'\to-S'\) parity.
\begin{equation}
 V=\lambda_S|S |^4+\frac{\lambda_{SS'}}2|S |^2{S'}^2
                  +\frac{\lambda_{S'}}{4}{S'}^4,
 \qquad
 \mathscr{L}_{\rm Yuk}=-\frac{y_{S}}{\sqrt2}S \psi\psi+\mathrm{h.c.}
 \label{eq:complexmodel}
\end{equation}
The one-loop RGEs are
\begin{eqnsystem}{sys:RGE1}
(4\pi)^2\beta_{\lambda_S}&=&20\lambda_S^2+\tfrac12\lambda_{SS'}^2+4\lambda_S y_{S}^2-4y_{S}^4,\\
(4\pi)^2\beta_{\lambda_{SS'}}&=&4\lambda_{SS'}^2+8\lambda_S\lambda_{SS'}+6\lambda_{SS'}\lambda_{S'}+2\lambda_{SS'} y_{S}^2,\\
(4\pi)^2\beta_{\lambda_{S'}}&=&18\lambda_{S'}^2+\lambda_{SS'}^2,\\
(4\pi)^2\beta_{y_S}&=&3y_{S}^3.
\end{eqnsystem}
We focus on the simpler minimum where only the charged $S$ acquires a vacuum expectation value.\footnote{Little changes in the opposite case where the neutral $S'$ gets involved too: 
$S'$ acquires thermal masses but has no chemical potential.
The potential has two minima at $\pm S'$;
the resulting domain walls  are cosmologically excluded unless 
the $S'\to - S'$ parity is explicitly broken.}
The radial potential again takes the form of eq.\eq{VSlog2}, with
\begin{equation}
y_{S}^2 = \frac{\lambda_{SS'}}{2\sqrt2},\qquad
 b=\frac{\lambda_{SS'}^2[6\lambda_{S'}+(4-\sqrt2)\lambda_{SS'}]}{8(4\pi)^4}. 
 \label{eq:complexb}
\end{equation}
The `dark' charge can be carried both by particles in the thermal bath with density $n_{\rm bath}$ and by a spatially homogeneous scalar condensate with density $n_0=s^2_0\dot\theta_0$~\cite{1910.02080,1910.14152,2108.09299}.
The total density $n_Q = n_0 + n_{\rm bath}$ of the conserved charge dilutes following the universe expansion, $ \dot n_Q+3Hn_Q=0$. The energy density of the condensate is given by
\beq
 \rho_0=\frac{\dot s_0^2}{2}+\frac{n_0^2}{2s_0^2}+V_0(s_0).
 \label{eq:cond_energy}
\eeq
We can assume the quasi-static approximation, in which the radial kinetic energy is neglected.
The divergence at $s_0=0$ of the `centrifugal potential'
means that a vanishing condensate cannot carry charge,
and we need to compute how charge is shared between
the condensate and the bath before minimising over the condensate radius.
Our goal is the free energy $F(s_0,T,n_Q)$ at fixed charge, whose minimum over $s_0$ determines the condensate. Charge sharing is most simply handled through grand potentials: in chemical equilibrium each component with charge $q$ has chemical potential $q\mu$, where a single $\mu$ is conjugate to the conserved charge, so their grand potentials just add.
We compute them in section~\ref{sec:Omega}.
Next $F$ follows by a Legendre transform in section~\ref{sec:F},
and is later used to study the cosmology.

\subsection{The grand potential}\label{sec:Omega}
The chemical potential of the condensate is its energy cost per unit charge at fixed $s_0$, so $\mu = \partial \rho_0/\partial n_0 = \dot\theta_0$ and the grand potential density of the condensate $\Omega_0 = \rho_0 - \mu n_0$ is
\begin{equation}
\Omega_0(s_0,\mu) = V_0(s_0) - \frac{\mu^2}{2} s_0^2,\qquad  n_0 = -\frac{\partial\Omega_0}{\partial\mu} = \mu s_0^2.
\end{equation}
We assume sufficiently fast charge exchange to establish chemical equilibrium among the various components.
The scalar and its condensate have chemical potential $\mu_s=\mu$ and the Weyl fermion has chemical potential $\mu_\psi=-\mu/2$.
The total grand potential density is $\Omega=\Omega_0+\Omega_{\rm bath}$. 
The thermal bath contribution is
\begin{equation}\label{eq:VT}
 \Omega_{\rm bath}(s_0,T,\mu)=\frac{T^4}{2\pi^2}
 \left[\sum_b n_bJ_B\!\left(\frac{m_b^2	}{T^2},\frac{\mu_b}{T}\right)
      +\sum_f n_fJ_F\!\left(\frac{m_f^2}{T^2},\frac{\mu_f}{T}\right)\right],
\end{equation}
where $m_{b/f}$ are background-dependent physical masses and the thermal integrals $J_B$ and $J_F$ are\footnote{Their small-mass expansions are
\begin{align}
 J_B(z,\xi)&=-\frac{\pi^4}{45}+\frac{\pi^2(z-2\xi^2)}{12}
          -\frac\pi6(z-\xi^2)^{3/2}-\frac{z^2}{32}\ln\frac{z}{a_B}+\cdots,\\
 J_F(z,\xi)&=-\frac{7\pi^4}{360}+\frac{\pi^2(z-2\xi^2)}{24}
          -\frac{\xi^4}{24}+\frac{z^2}{32}\ln\frac{z}{a_F}+\cdots,
\end{align}
where  $a_B=16a_F=16\pi^2e^{3/2-2\gamma_E}$.
At small background field the logarithms combine with the
vacuum contribution, so the relevant running couplings are evaluated near $\bar\mu\sim T$.
Mixed $z\xi^2$ terms in $J_F$, which depend on the renormalisation of the $S$ kinetic term, are neglected.
}
\begin{equation}
	J_{B / F}\left(y^2 ,\xi\right)= \frac{P}{2} \int_0^{\infty} dx\, x^2 \sum_\pm
	\ln \left[1 - P \exp \left(-\sqrt{x^2+y^2} \pm \xi\right)\right],
\end{equation}
with $P=+1$ for bosons and $P=-1$ for fermions. 
The two signs correspond to particles and antiparticles.
This expression holds for fermions with charge-conserving Dirac masses.
A different $J_F$ applies to our Weyl fermion $\psi$, whose Majorana mass violates $\psi$ number:
the quasi-particle energies become $E_\pm=\sqrt{(p\pm\mu_\psi)^2+m_\psi^2}$, giving
\begin{equation}
	J_F^{\rm Weyl}(y^2,\xi)= -\frac12\int_0^\infty dx\,x^2\bigg[\sum_\pm \ln\!\Big(2\cosh\frac{\sqrt{(x\pm\xi)^2+y^2}}{2}\Big)-\sqrt{x^2+y^2}\bigg]_{\rm ren},
\end{equation}
where $\sqrt{x^2+y^2}$ removes the zero-temperature term already in $V_0$ and the leftover divergence $\propto y^2\xi^2$ renormalises the $S$ kinetic term. We only need the cases $m_\psi=0$ or $\mu_\psi=0$, where $J_F^{\rm Weyl}=J_F$.

The radial $\delta s$ and angular $\delta\theta$ fluctuations of the complex scalar $S$  only have radiatively generated masses, which we neglect, treating both modes as relativistic. In contrast, the field-dependent masses of $S'$ and $\psi$ are
\begin{equation}
 m_{S'}^2(s_0)= \sqrt{2} y^2_Ss^2_0  = \frac{\lambda_{SS'}}{2}s_0^2,
  \qquad
 m^2_\psi(s_0)= y^2_Ss^2_0 = \frac{\lambda_{SS'}}{2\sqrt{2}}s_0^2 ,
\end{equation}
where the criticality condition of eq.\eq{complexb} was used. 

\subsubsection*{Grand potential at large field values}
The thermal potential saturates at field values beyond a critical $s_c = T/y_S$, where both $\psi$ and $S'$ become heavy.
In this limit
\begin{equation}
 \Omega_{\rm bath}(s_0,T,\mu) \simeq
-\frac{19\pi^2}{360}T^4+AT^4
-\frac12 c_\chi\mu^2T^2,
\qquad n_{\rm bath} = -\frac{\partial\Omega}{\partial\mu} = c_\chi \mu T^2,
\label{eq:Omega_bathNR}
\end{equation}
where a susceptibility coefficient $c_\chi=1/3$ includes only contributions from $S$.
The first term is the free energy of the full massless plasma; the plasma has however lost the $\Delta g=1+\frac78\cdot2=\frac{11}4$ degrees of freedom of $S'$ and $\psi$, which costs $AT^4$ with $A=\pi^2\Delta g/90=11\pi^2/360$.
The total grand potential density in this regime is
\begin{equation}
 \Omega(s_0,T,\mu) =  -\frac{19\pi^2}{360}T^4 + AT^4+ V_0(s_0) - \frac12 \mu^2 (s_0^2+c_\chi T^2).
 \label{eq:OmegaNR}
\end{equation}

\subsubsection*{Grand potential at small field values}
In the opposite regime, when $s_0\ll s_c$ and all masses remain small compared to the temperature $m_{b/f}\ll T$ we can expand
\begin{equation}
\begin{split}
\Omega_{\rm bath}(s_0,T,\mu) \simeq
&-\frac{19\pi^2}{360}T^4
-\frac12\tilde{c}_\chi\mu^2T^2
+\frac12 M_T^2 s_0^2\\
&-\frac{T}{12\pi}\Bigg[2\left(
M_T^2-\mu^2
\right)^{3/2}
+
\left(
M_{T}^{'2}+\frac{\lambda_{SS'}}{2}s_0^2
\right)^{3/2}
\Bigg]  + \mathcal{O}(m_{b/f}^4),
\label{eq:Omega_bath}
\end{split}
\end{equation}
where in the non-analytic bosonic $m_b^3$ terms
the masses of $S$ and $S'$ were replaced by their ring-resummed values \cite{hep-ph/9212235,1701.01554}.
The coefficient $\tilde{c}_\chi=3/8$ now also includes the fermionic contribution
and the thermal masses for $S$ and $S'$ are
\begin{equation}
 M_T^2 = c_T^2 T^2,\quad c_T^2
=
\frac{1+\sqrt{2}}{24\sqrt{2}}\lambda_{SS'}
,\quad\quad
M_{T}^{'2}
=
\left(
\frac{\lambda_{S'}}{4}
+\frac{\lambda_{SS'}}{12}
\right)T^2.
\end{equation}
Neglecting the ring terms and higher orders in $m_{b/f}^2/T^2$, the total grand potential is
\begin{equation}
 \Omega(s_0,T,\mu) \simeq V_0(s_0) -\frac{19\pi^2}{360}T^4 - \frac{1}{2}\tilde{c}_\chi \mu^2 T^2 + \frac12 (M_T^2-\mu^2)s_0^2 \qquad (s_0\ll s_c).
 \label{eq:OmegaR}
\end{equation}

\subsection{The free energy}\label{sec:F}
Since charge conservation in the expanding universe fixes the charge density $n_Q$, rather than $\mu$, the equilibrium value of $s_0$ is found by minimising the free energy $F$ at fixed $T$ and $n_Q$.
It is obtained by the Legendre transformation $F=\Omega+\mu n_Q$, with $\mu$ determined by $n_Q=-\partial\Omega/\partial\mu$.
Applied to eq.s\eq{OmegaR} and\eq{OmegaNR}, it gives the free energy in the two regimes:
\begin{equation}
 F(s_0,T,n_Q) - F(0,T,0)\simeq V_0(s_0)+
 \begin{cases}
\displaystyle\frac{n_Q^2}{2(s_0^2+\tilde{c}_\chi T^2)} +  \frac{c_T^2}{2}  T^2 s_0^2   , & s_0 < s_*\\
  \displaystyle \frac{n_Q^2}{2(s_0^2+c_\chi T^2)}+A T^4     & s_0 > s_*
 \end{cases},
\label{eq:free_energyRandNR}
\end{equation}
where $F(0,T,0)=-19\pi^2T^4/360$.
The two limits are matched at an intermediate $s_*$ such that $F$ is continuous.
Approximating $c_\chi=\tilde{c}_\chi$ gives $s_* \approx 1.73\, s_c$.
At $s_0>s_*$ the condensate carries almost all the charge, so the bath contribution $c_\chi T^2$ can be neglected.
The free energy remains finite at the origin because the bath can carry some charge.

At small field values, neglecting $V_0$, the origin is a maximum of $F$ if $n_Q>\tilde c_\chi c_T T^3$, and the minimum is at
\begin{equation}
 \mu = M_T,\qquad n_0 = n_Q - n_{\rm bath},\qquad  n_{\rm bath}= \tilde{c}_\chi c_T T^3,
\qquad s_0^2 = \left(\frac{n_Q}{c_T T^3}- \tilde{c}_\chi\right) T^2 ,
\label{eq:equilibrium}
\end{equation}
where $\mu=n_Q/(s_0^2+\tilde c_\chi T^2)$. Neglecting $V_0$ is justified by the loop suppression $b\ll 1$ and by $s_0\ll s_c$\footnote{This term becomes relevant only at exponentially large temperatures due to the logarithmic running of~$\lambda_S$.}.
When the condensate forms, the charge stored in the thermal bath is saturated, and the excess goes into the condensate.
The leading corrections to the equilibrium relation for $\mu$ and $s_0$, from the $S'$ ring term in eq.\eq{Omega_bath}, remain under control for perturbative quartic couplings.

\begin{figure}[t]
$$\includegraphics[width=0.45\textwidth]{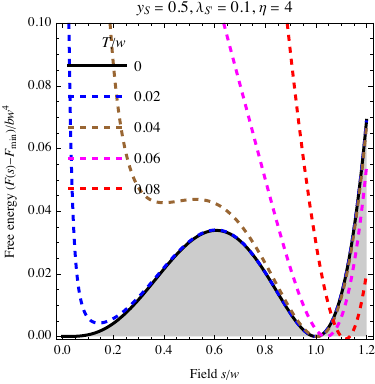}\qquad
\includegraphics[width=0.45\textwidth]{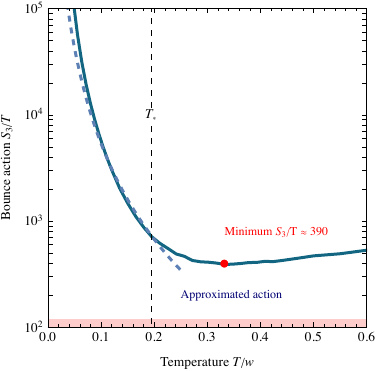}$$
\begin{center}
\caption{\em\label{fig:HonestColemanWeinbergVrot} Left: free energy potential in the 
model of eq.\eq{complexmodel} with the indicated parameters.
Right: bounce action $S_3/T$ of eq.\eq{bounce_action} for nucleating symmetric bubbles, compared with the estimate of eq.\eq{gradientbound}.}
\end{center}
\end{figure}

\subsection{Formation of the broken phase}
We now study the cosmological evolution and determine under which conditions
it selects the broken vacuum.
If the total charge and the entropy are conserved, the charge yield $Y_Q\equiv n_Q/s_{\rm ent}$ remains fixed: over a temperature range in which the number of entropy degrees of freedom does not change significantly, $n_Q=\eta T^3$ with approximately constant $\eta$.

If $\eta>\tilde{c}_\chi c_T$ a condensate forms: the free energy in the $s_0< s_*$ regime has a local maximum at $s_0=0$ and a minimum at the $s_0\propto T$ of eq.\eq{equilibrium}, neglecting $V_0$ as before. 
Since both $n_Q^{1/3}$ and $M_T$ scale as $T$, this criterion is temperature independent: either the condensate forms as soon as the charge is present, or never. The condensate therefore cannot form later from a charged plasma: it must exist since the charge is generated. We assume it is produced coherently, as in the Affleck--Dine mechanism discussed below, so that inflation makes it homogeneous. If instead the charge were generated in the plasma, the condensate would grow from the instability of the origin with random phases, forming strings and inhomogeneities whose relaxation is beyond the scope of this work. 
For smaller $\eta$, such a condensate tracks $s_0\propto T$ towards the origin. The broken vacuum can only be selected for
\begin{equation}
 \eta>c_T\left(\tilde{c}_\chi + \frac{s_*^2}{T^2}\right) \approx \frac{1.35}{y_S},
 \label{eq:etaland}
\end{equation}
such that the minimum lies in the $s_0>s_*$ range, where the thermal force is switched off, and is determined by the balance between the outward centrifugal force and the inward potential force:
\begin{equation}
 V_0'(s_0)
    = 2b s_0^3 \ln\left(\frac{s_0}{w}\right)
    \left[2\ln\left(\frac{s_0}{w}\right)+1\right]
    = \frac{\eta^2 T^6}{s_0^3}.
\end{equation}
Eq.\eq{etaland} corresponds to $Y_Q\gtrsim 0.03/y_S$ for the SM number of degrees of freedom, $g_{*s}\approx106.75$.
At sufficiently high temperatures the condensate sits at $s_0\gg w$, and moves towards the minimum at $s_0=w$ as the temperature decreases. If it reaches the barrier while $s_0<s_*$, the thermal mass overcomes the barrier and pushes the field towards the origin; otherwise the field is stopped by the barrier at $s_0=w$. In this case an inner minimum also arises at $s_{\rm in}\gtrsim s_*$, but its free energy is higher because of the larger centrifugal term and $V_0(s_{\rm in})\neq 0$. 
At $T\lesssim m_{\delta s}(w)=\sqrt{2b}\,w$ the radial mode decouples at $s_0=w$ but not at the inner minimum, which thereby gains at most $\Delta F_{\delta s}\sim\pi^2T^4/90\sim b^2w^4$.
This is too small to compensate the larger centrifugal energy at $s_{\rm in}$ for $\eta$ satisfying eq.\eq{etaland}, and in any case parametrically smaller than the barrier $\sim bw^4$: thermal nucleation is suppressed and the condensate remains trapped in the outer minimum.
Fig.\fig{HonestColemanWeinbergVrot} illustrates an example,
with couplings that run perturbatively up to the Planck scale.
This remains true if an extra interaction $\lambda_{HS}|S|^2|H|^2$ with small $|\lambda_{HS}|$ triggers electroweak symmetry breaking at lower temperature.

It can also happen that $s_0$ is initially close to the origin and the temperature is never high enough to push the condensate over the barrier.
Since $V_0+n_Q^2/2s_0^2$ has an inner minimum unless $n_Q>0.0918\sqrt b\,w^3$,
avoiding it requires
\begin{equation}\label{eq:Ti}
 T_{\rm start}\gtrsim w\left(\frac{0.09\sqrt b}{\eta}\right)^{1/3}.
\end{equation}
This condition can be relaxed if the field starts in a local minimum
and reaches the global minimum through a first-order phase transition
(e.g.\ the brown potential in fig.\fig{HonestColemanWeinbergVrot}).

The required large charge asymmetry is naturally generated by an Affleck--Dine-like
mechanism~\cite{Affleck:1984fy,hep-ph/9507453}: U(1)-breaking
higher-dimensional operators exert a torque on an initially displaced
scalar, generating a coherent rotation; their effects become
negligible as the field amplitude decreases, leaving an approximately
conserved comoving charge. Applications to rotating axions are
discussed in~\cite{1910.02080,1910.14152}.

More complicated scenarios are possible under special conditions.
If the total $M^2_\phi$ changes sign (e.g.\ because of running couplings)
and becomes negative only below some temperature, a first-order phase transition may reach the broken vacuum.
If the U(1) is global and exact, the massless Goldstone boson acts as a small amount of extra radiation.
If the angular mode $\theta$ has a small U(1)-breaking potential,
its torque averages out over each revolution, until $\theta$ gets trapped at its minimum.
The global U(1) of the minimal model has a mixed gravitational anomaly:
its contribution to charge non-conservation is proportional
to $R\widetilde R$, which vanishes in a cosmological background.
The anomaly can be cancelled by adding another fermion,
after which the U(1) can be gauged.
A gauged U(1) must satisfy Gauss's law,
so extra fields are needed to cancel the total charge,
while interactions should respect a global U(1) that allows for separate asymmetries.

%\subsection{Survival of the broken phase}\label{sec:survival}
The previous calculations assumed homogeneity.
Selecting the broken vacuum
throughout space requires two further checks: whether charge prefers to
concentrate in only part of the volume, and whether an already formed
homogeneous broken phase survives subsequent cooling. These are distinct
thermodynamic and dynamical questions. They arise in the thermodynamics
of $Q$-balls~\cite{hep-ph/9804237} and in cosmological models of rotating
fields coupled to a thermal bath~\cite{2205.00942}.

\subsection{Does charge favour a homogeneous broken phase?}\label{sec:coexist}
The broken phase stores charge cheaply in coherent rotation, but pays a
thermal free-energy cost because some particles become heavy. The two phases can exchange charge, so it can be favourable to concentrate part of the total charge in broken regions while leaving the rest of space symmetric.
In equilibrium, they have a common chemical potential $\mu_c$ and equal pressures $p=-\Omega$.
Let $s_b$ be the condensate radius in the broken phase. The two conditions are: the rotating condensate is stationary, with the centrifugal force balancing $V_0'$; the broken phase, eq.\eq{OmegaNR} at $s_0=s_b$, and the symmetric phase, eq.\eq{OmegaR} at $s_0=0$, have equal pressures:
\begin{equation}
 V_0'(s_b)=\mu_c^2s_b,\qquad
 AT^4+V_0(s_b)=\frac{\mu_c^2}{2}
 \left[s_b^2+(c_\chi-\tilde{c}_\chi)T^2\right].
 \label{eq:coexist_pressures}
\end{equation}
The symmetric phase must also remain locally stable,
$\mu_c^2<c_T^2T^2$.  Neglecting interfaces, a mixture with a fraction $f$ of the volume in the broken phase shares the total charge between the symmetric and broken regions, with charge densities $n_s$ and $n_b$:
\begin{equation}
 \eta T^3=(1-f)n_s+fn_b,\qquad n_s=\tilde c_\chi\mu_cT^2,\qquad n_b=\mu_c(s_b^2+c_\chi T^2).
 \label{eq:mixture}
\end{equation}
Eq.\eq{coexist_pressures} determines $s_b$ and $\mu_c$; 
charge conservation in eq.\eq{mixture}
then fixes the equilibrium fraction of the volume in the broken phase, $f_{\rm eq}=(\eta T^3-n_s)/(n_b-n_s)$.
For $\epsilon\equiv4A^3/b\eta^4\ll1$ the broken phase sits close to the potential minimum, $s_b\simeq w$, so that $\mu_c\simeq\sqrt{2A}\,T^2/w$ and $n_s\ll n_b\simeq\mu_c w^2$. This gives
\begin{equation}
 f_{\rm eq}\simeq\frac{T}{T_b},\qquad T_b=\frac{\sqrt{2A}}{\eta}\,w.
 \label{eq:fullvolume}
\end{equation}
The homogeneous broken phase is thermodynamically preferred for $T\geq T_b$, while at lower temperatures only a fraction $T/T_b$ of the volume would be broken.
Whenever the broken phase forms, i.e.\ for $\eta$ satisfying eq.\eq{etaland}, the symmetric phase is locally stable ($\mu_c<c_TT$ requires $\eta>2A/c_T$), so below $T_b$ a mixture of the two phases has lower free energy than the homogeneous broken phase. The homogeneous broken phase is then metastable: whether it actually fragments depends on the nucleation rate of symmetric bubbles, computed in section~\ref{sec:retention}. 
Larger $\eta$ helps, as it lowers $T_b$. For very large $\eta$, $T_b$ falls below the radial mass $m_{\delta s}(w)=\sqrt{2b}\,w$: the radial mode then is also heavy in the broken phase, which only increases $A$ by $\pi^2/90$ and $T_b$ by about 17\%.

\subsection{Does the broken phase survive cooling?}\label{sec:retention}
The next question is whether the thermodynamic preference for $f_{\rm eq}<1$ leads to fast fragmentation of the homogeneous broken phase. The condensate approaches $s_0=w$ during cooling but it may become metastable against nucleation of symmetric phase bubbles when a mixed state becomes preferred.

Linearising the field equations around a circular homogeneous orbit, 
$s=s_0 +\delta s$, $\theta=\omega t +\delta\theta$, shows that
the rotating condensate is locally stable if its radial squared mass is positive: 
at low temperature $V_0''-\omega^2=2bs_0^2[1+6\ln(s_0/w)+4\ln^2(s_0/w)]>0$ for $s_0\gtrsim w$, while near $T_b$ thermal corrections can spoil this, so it is checked numerically below. 

Without a local instability, conversion requires bubble nucleation.
The nucleation rate is given by $\Gamma_{\rm nuc}\sim T^4\, e^{-S_3/T}$ where $S_3$ is a bounce action.
A conversion rate slower than Hubble $H^4$ needs  $S_3/T \gtrsim 4\ln T/H \sim 120$ at $T \sim 100\TeV$.
The bounce action can be approximated as follows:\footnote{The bounce action can also be analytically computed for $\epsilon\ll1$, where it is very large.
The pressure driving a small symmetric region in a homogeneous broken reservoir is
$\Delta p=\Omega_b-\Omega_s\simeq AT^4[1-(T/T_b)^2]$; its maximum is smaller than the barrier $V_{\rm bar}=bw^4/4e^2$ by a factor $\simeq4.38\,\epsilon$,
so the potential is well approximated by $V_0$ and the bounce is a thin wall with action
$S_3/T\simeq16\pi\sigma^3/3T\Delta p^2$, where $\sigma\simeq\int_0^w ds\sqrt{2V_0(s)}=\sqrt{2b}\, w^3/9$ is the wall tension.
Its minimum over $T<T_b$ is $(S_3/T)_{\min}\simeq384\,b^{3/2}\eta^9$.}
a symmetric bubble must displace the field
by $ {\cal O}(w)$ over its radius, so its gradient energy is at least $2\pi {\cal O}(w)^2R$, where $R$ is the radius of the
ball with the same volume, while the gain is at most $\frac{4\pi}3R^3AT^4$. Maximising over $R$ gives
\begin{equation}
 \frac{S_3}{T} \approx \frac{4\pi}{3\sqrt{2A}}\left(\frac{ {\cal O}(w)}{T}\right)^3\sim 5.4\left(\frac {{\cal O}(w)}{T}\right)^3 .
 \label{eq:gradientbound}
\end{equation}
Inserting the maximal $T_b$ suggests that nucleation is slow for $\eta \gtrsim 2$.
We recall that $\eta > 1.35/y_S$ is anyhow needed by eq.\eq{etaland}.

\medskip

To establish viability more precisely, we numerically compute the
action $S_3$ of the spherically symmetric thermal bounce $s(r)$.
During nucleation the bubble can exchange charge with the surrounding condensate, which acts as a charge reservoir:
the chemical potential stays fixed at its background value $\mu=\eta T^3/s_0^2$,
while the local charge density adjusts as $n_Q(r)=-\partial\Omega/\partial\mu$, consistently with conservation of the total charge.
The bounce is therefore computed from the grand potential $\Omega(s,T,\mu)$ at fixed $\mu$.
The symmetric phase inside the bubble must be locally stable, $\mu<c_TT$, which also ensures the validity of the small $\mu/T$ expansion.
When the symmetric phase has a lower grand potential, the bounce satisfies
\beq
 \frac{d^2s}{dr^2}+\frac{2}{r}\frac{ds}{dr}
 =\frac{\partial\Omega(s,T,\mu)}{\partial s},
 \qquad
 s'(0)=0,\qquad s(\infty)=s_0,
 \label{eq:bounce_equation}
\eeq
and its action is
\beq
 \frac{S_3}{T}=\frac{4\pi}{T}\int_0^\infty dr\,r^2
 \left[
 \frac12\left(\frac{ds}{dr}\right)^2
 +\Omega(s,T,\mu)-\Omega(s_0,T,\mu)
 \right].
 \label{eq:bounce_action}
\eeq
The right panel of fig.\fig{HonestColemanWeinbergVrot} shows $S_3/T$
as a function of the temperature, for model parameters such that nucleation is slow enough.
We see that eq.\eq{gradientbound} provides a reasonable approximation for $S_3/T$ at $T < T_b$.
The plot extends above $T_b$ because eq.\eq{fullvolume} assumed $s_0=w$, while the numerical computation does not.
It verifies that the broken phase is long-lived, both against local instabilities and against bubble nucleation,
for $\eta \gtrsim 3.8$.

If nucleation occurs, charge accumulates in the remaining broken regions and halts the conversion before it completes.
As the universe cools, the resulting $Q$-balls expand, but the universe expands faster,
so that the symmetric phase ends up occupying almost all the volume.
See also~\cite{Griest:1989bq,hep-ph/9804237,2205.00942,2512.16601,2206.14259}.

Cosmic strings, if present, could enhance the nucleation rate~\cite{2510.27579}, since their core
is a pre-existing region of symmetric phase, comparable in size to the critical bubble over the relevant temperature range.
However, no strings form for a coherent Affleck--Dine condensate, since the U(1) is never restored: the origin is unstable at all temperatures.

\section{Conclusions}\label{concl}
We studied cosmological vacuum selection in theories where dimensional transmutation generates a symmetry-breaking minimum tuned to have zero vacuum energy, degenerate with the symmetric vacuum. 

We found that a neutral thermal plasma favours the symmetric phase because the leading thermal mass squared is positive along the nearly flat ray in field space that contains the broken vacuum. 
Negative mixed quartics can induce negative squared masses and destabilise the origin in other directions, 
but this does not lead to cosmological histories that select the broken vacuum.

A sufficiently large conserved charge asymmetry can overcome this obstruction by supporting a rotating scalar condensate that approaches the broken vacuum as the charge dilutes. Successful selection requires both reaching this branch and preserving the broken phase against conversion into a mixed state. Although the cold outer condensate is linearly stable, thermal nucleation and charge redistribution can affect its survival. 
The condition for the condensate to reach the broken vacuum is given in eq.\eq{etaland}; 
the extra condition for the resulting phase to be locally stable and long-lived is discussed in sections~\ref{sec:coexist} and~\ref{sec:retention}.
Together they require a dark charge asymmetry $n_Q/s\gtrsim 0.05$, orders of magnitude larger than the baryonic one.

Under these conditions, the mechanism therefore provides a possible cosmological route to the broken vacuum.
Unlike the usual Coleman--Weinberg scenarios, the selection can minimally proceed without a first-order
transition, so no gravitational waves from bubble collisions are produced.

\paragraph*{Acknowledgements}
AS thanks Claude and ChatGPT for useful discussions.

\appendix

\end{document}